\documentclass[11pt,a4paper]{article}
\usepackage{graphicx}
\usepackage{amsfonts}
\usepackage{amssymb}
\def\RE{\ensuremath{\mathop\mathrm{Re}\nolimits}}
\def\RT{\ensuremath{\RE_{\rm t}}}
\def\RG{\ensuremath{\RE_{\rm g}}}

\def\BF{\begin{figure}}
\def\EF{\end{figure}}
\def\BC{\begin{center}}
\def\EC{\end{center}}

\def\BE{\begin{equation}}
\def\EE{\end{equation}}
\def\rmd{\mathrm{d}}
\def\BI{\begin{itemize}}
\def\EI{\end{itemize}}

\title{On the growth of laminar-turbulent patterns in plane Couette flow}

\author{Paul Manneville\\Hydrodynamics Laboratory\\
\'Ecole Polytechnique,
Palaiseau 91128, France\\
\tt paul.manneville@polytechnique.edu}
\date{to appear in Fluid Dyn. Res. (BIFD 11, special issue)}

\begin{document}
\maketitle

\begin{abstract}
The growth of laminar-turbulent band patterns in plane Couette flow is studied in the vicinity of the global stability threshold \RG\ below which laminar flow ultimately prevails.
Appropriately tailored direct numerical simulations are performed to manage systems extended enough to accommodate several bands.
The initial state or {\it germ\/} is an oblique turbulent patch of limited extent.
The growth is seen to result from several competing processes: ({\it i\/}) nucleation of turbulent patches close to or at the extremities of already formed band segments, with the same obliquity as the germ or the opposite one, and ({\it ii\/}) turbulence collapse similar to gap formation for band decay.
Growth into a labyrinthine pattern is observed as soon as spanwise expansion is effective.
An ideally aligned pattern is usually obtained at the end of a long and gradual regularisation stage when \RE\ is large enough.
Stable isolated bands can be observed slightly above \RG.
When growth rates are not large enough, the germ decays at the end of a long transient, similar to what was observed in experiments.
Local continuous growth/decay {\it microscopic\/} mechanisms are seen to compete with large deviations which are the cause of {\it mesoscopic\/} nucleation events (turbulent patches or laminar gaps) controlling the {\it macroscopic\/} behaviour of the system (pattern).
Implications of these findings are discussed in the light of Pomeau's proposals based on directed percolation and first-order phase transitions in statistical physics.

\end{abstract}

\vspace{2pc}

\maketitle
\sloppy

\section{Context\label{sec1}}

The direct transition to turbulence in wall-bounded flows still raises open questions linked to the metastability of turbulence and the possible coexistence of laminar and turbulent flow in a finite range of Reynolds numbers. 
Important progress has been achieved recently for pipe flow in a cylindrical tube where a well-defined threshold could be defined by comparing rates for {\it puff decay\/} and {\it puff splitting\/}:
Splitting propagates turbulence while decay drives the flow to the laminar state, so that turbulence can persist indefinitely with finite probability when new puffs are produced faster than they collapse (Moxey \& Barkley, 2010; Avila {\it et al} 2011; Barkley 2011).
Plane Couette flow, the flow between counter-translating parallel plates, represents another canonical flow situation with a direct transition observed while laminar flow is linearly stable for all Reynolds numbers.
With \RE\ being defined as $Uh/\nu$, where $2U$ is the relative speed of the plates, $2h$ the gap between them,  and $\nu$ the kinematic viscosity of the sheared fluid, two thresholds have been identified.
Laminar flow is always recovered in the long time limit below the lower threshold \RG.
Experiments performed by the Saclay group (Bottin 1998; Bottin {\it et al.} 1998) situated it at  $\RG\approx325$.
Featureless turbulence is obtained above the upper threshold \RT\ and, between \RG\ and \RT, laminar and turbulent flow coexist in space.
Provided that the aspect-ratio -- the ratio of lateral setup dimensions to the gap -- is sufficiently large, this coexistence manifests itself in the form of oblique bands, alternately laminar and turbulent, at rest in the laboratory frame (Prigent 2001; Prigent {\it et al.} 2003).
The Saclay experiments have shown that $\RT\approx410$ and, in units of $h$ as used everywhere in the following, the streamwise wavelength of the turbulence modulation is $\lambda_x\approx110$, while the spanwise wavelength $\lambda_z$  varies from 55--65 close to \RT\ to 70--90 around \RG.

A previous study  (Manneville 2011) was dedicated to the decay of these bands for $\RE\lesssim\RG$, i.e. the {\it turbulent$\>\to\>$laminar\/}  transition.
Here we examine the {\it laminar$\>\to\>$turbulent\/} transition  for $\RE\gtrsim\RG$, not the early stage where a localised sufficiently strong perturbation is turned into a turbulent spot but rather the late stage, i.e. the growth of a developed pattern from a small but sufficiently extended turbulent patch.
Turbulent spot dynamics in simple shear flow has been studied first by Lundbladh \& Johansson (1991) numerically in computer domains of size $128\times2\times64$ at Reynolds numbers somewhat larger than those we are interested in.
Accordingly, their spots grew quickly while mostly keeping their initial ovoid shape. Laboratory experiments were later performed in the same regime by Dauchot \& Daviaud (1995) and Tillmark (1995).
On the other hand, Bottin (1998) studied the behaviour of spots in the vicinity of \RG, the determination of which she largely contributed to [see also (Bottin {\it et al} 1998)].
The spanwise dimension of her set-up ($L_x=284$, $L_z=72$) was however insufficient to show the bands and only her very latest experiments at doubled aspect ratio could show a tendency to form oblique patches at steady state.
In contrast, bands were conspicuous in Prigent's work with $L_x=770$ and $L_z=340$ (Prigent 2001; Prigent {\it et al.} 2003) but these researchers studied only patterning well above \RG, closer to \RT, by gradually decreasing \RE.

Direct numerical simulations (DNS) of the Navier--Stokes equations in domains sufficiently extended to be of interest to pattern formation have been performed by Duguet {\it et al.} (2010) who considered the evolution of the flow from a sufficiently intense random noise  in a domain of size comparable to Prigent's ($800\times2\times356$): 
The many spots left after a brief viscous smoothing stage were seen to interact and generate a kind of {\it labyrinthine structure\/} that became gradually organised into a  patchwork of well oriented domains separated by grain boundaries (see movie \#2 attached to their article).
Motivated by the work done in pipe flow, Duguet {\it et al.} (2011) next focused on the statistics of the laminar-turbulent interface in a spanwise-elongated but streamwise-narrow computational domain ($10\times2\times250$), i.e. a confined quasi-one dimensional configuration similar to a pipe.
They pointed out a competition between {\it progress events\/} (sudden expansion of the turbulent domain, reminiscent of puff splitting) overcoming {\it retreat events\/} (sudden breakdown of turbulence, reminiscent of puff decay) beyond some threshold located in the vicinity of \RG\ but possibly slightly different. 
Their elongated domain however did not allow the development of an average large scale laminar flow outside the turbulent domain induced by Reynolds stresses inside it, as pointed out by e.g. Lundbladh \& Johansson (1991) in their DNS, by Dauchot \& Daviaud (1995) or Tillmark (1995) in their experiments, by Schumacher \& Eckhardt (2001) using DNS with artificial stress-free boundary conditions, by Barkley \& Tuckerman (2007, 2011) using their oblique narrow domain, or by Lagha \& Manneville (2007b) using a model.

In recent years, we have been involved in a numerical study of plane Couette flow dedicated to large aspect ratio systems free of limitations implied by confinement in the streamwise direction (Duguet {\it et al.} 2011) or skewed streamwise direction (Barkley \& Tuckerman 2005).
Aiming at the same goal as Duguet {\it et al.} (2010) but at much reduced computational cost, we validated a modelling strategy based on DNS at controlled under-resolution (Manneville \& Rolland 2010).
The subsequent sections expand (\S\ref{sec2}) and discuss (\S\ref{sec3}) our results focused on the growth of patterns.
We refer to (Manneville 2011) for further details about the computing methodology.
In that article, decay was shown to result from a small-scale stochastic process generated by chaos in the range of Reynolds numbers where turbulent bands are present.
The transient nature of this chaos makes the decay of turbulent bands toward laminar flow possible.
A competition, however, exists between local decay and contamination of locally laminar flow by turbulence at the laminar-turbulent interface.

As conjectured by Pomeau (1986), {\it directed percolation\/} -- as involved in e.g. flow through porous media, forest fires, or propagation of epidemics-- becomes a relevant framework (Hinrichsen 2000) to interpret the transition from expanding to retracting turbulence at the extremities of an oblique band segment. In order to operate, this contamination process first demands  that the continuous band be broken by the opening of a sufficiently wide laminar gap. In turn, the opening of such a gap results from {\it large deviations\/} of the small-scale transient chaotic dynamics. 
Here, growth from a germ (a localised patch of turbulence) at $\RE\gtrsim \RG$ will be shown to involve the same directed percolation process, now biased towards growth instead of decay, and the same large-deviation effects. The latter will produce either the nucleation of a turbulent bud oriented in the direction complementary to that of the considered  oblique turbulent band segment, or the collapse of a large region taken in the turbulent patch, namely the opening of a laminar gap or the relaminarisation of a newly born turbulent bud.

\section{Results\label{sec2}}

\subsection{The growth experiment}

Simulations make use of \textsc{ChannelFlow}, the open-source software developed by Gibson (1999).
This software is a pseudo-spectral  Fourier~($x$) $\times$ Chebyshev~($y$) $\times$ Fourier~($z$) de-aliased scheme integrating the Navier--Stokes equations.
A good compromise between computational load and realism has been found for $N_x=L_x$ and $N_z=3L_z$, $N_{x,z}$ being the numbers of collocation points used in the evaluation of in-plane dependence of nonlinear terms, and the number of Chebyshev polynomials  $N_y=15$.
This resolution may seem quite low but all the features of the transitional regime appear to be well preserved (self-sustaining process, laminar-turbulent coexistence, pattern selection with $\lambda_x$ and $\lambda_z$ comparing favourably with experimental values).
The price to be paid is just a moderate shift of the transitional range $[R_{\rm g}, R_{\rm t}]$ by about 15\% from $[325,410]$ down to $[275,360]$.
A thorough validation of the numerical procedure can be found in  Manneville \& Rolland (2010).
On the other hand, we are able to study domains with typical size $L_x>400$, $L_z>250$, and perform many statistically significant experiments using merely the power of a desk-top computer.

The general conditions used to study the decay of bands (Manneville 2011) also hold for the growth from a germ studied here as it depends on the Reynolds number. The only difference is the size of the domain which has been slightly enlarged from (432,256) to (468,272) in an attempt to delay the effects of periodic boundary conditions as much as possible without greatly increasing the computational burden.
In the following, images will display the field of {\it local perturbation energy\/} averaged over the thickness:
\BE
e(x,z,t)=\frac12\int_{-1}^{+1}
\rmd y\,\left[\mbox{$\frac12$}\mathbf{\tilde v}^2\right],\label{e:et}
\EE
where $\mathbf{\tilde v}=\mathbf v - y\mathbf{\hat x}$, $\mathbf v$ being the total velocity field and $y\mathbf{\hat x}$ the laminar flow.
When necessary, we shall superpose the streamlines of the in-plane velocity perturbation field  $\bar u_x\mathbf{\hat x}+\bar u_z\mathbf{\hat z}$ averaged over the thickness:
\BE
\bar u_{x,z}(x,z,t)=\frac12\int_{-1}^{+1}
\rmd y\, u_{x,z}(x,y,z,t)\,,\label{e:uxz}
\EE
subsequently mildly filtered as explained in Manneville (2011) or display the local intensity of this flow as defined by $\sqrt{\bar u_x^2 + \bar u_z^2}$. On some occasions, taking advantage of in-plane periodic boundary conditions, we shall display a $2\times2$ matrix tiling of the solution to offer a better view of the pattern obtained.
Graphs will show the {\it distance to laminar flow\/} defined as
\BE
 \Delta(t)=\frac{1}{L_xL_z}\int\!\!\!\int\rmd x\,\rmd z\, \sqrt{2e(x,z,t)}\,.\label{e:met}
\EE

The same germ, shown in Fig.~\ref{f:germ}, has been used for all simulations, prepared from a small isolated turbulent oblique patch obtained at $\RE=278.75$ in the  (432,256) domain, and placed in the (468,272) domain.
The germ was extracted as a snapshot from a decay experiment, long before the latest stage of viscous relaxation, at a time when a single sizeable area of developed turbulence is still present, elongated and  obliquely oriented, with typical width/length of the order of a turbulent band width (Manneville 2011).
Starting from such a {\it mature spot\/} as the initial state has the advantage of skipping the early stage of growth from more limited but  strongly perturbed flow and to better focus on the selection processes at work during the formation of the turbulent band pattern.
Starting from highly localised initial states would work when $\RE\gg \RG$ and growth is essentially a deterministic process as studied by Lundbladh \& Johansson (1991) and Dauchot \& Daviaud (1995) but, for $\RE\gtrsim\RG$, most would decay. The very first growth stage indeed sensitively depends on the shape of early spots and the intensity of turbulence inside them, as shown by Bottin (1998).
On the other hand, starting from a random initial field like Duguet {\it et al.} (2010) would not produce sufficiently isolated spots (see their movie~\#2). Displaying streamlines of the mean in-plane flow as in Fig.~\ref{f:germ} might be misleading in that this overemphasises long-distance interactions while the flow is in fact exponentially weak away from the turbulent area. This has however the merit of reminding us of potential reconnection problems when several spots interact, a situation here mimicked by the periodic boundary conditions.  
\BF
\BC
\includegraphics[width=0.45\textwidth]{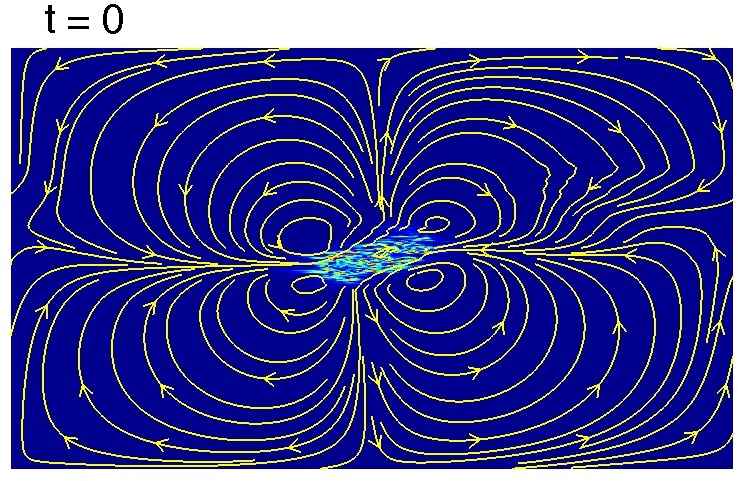}
\EC
\caption{Averaged local perturbation energy defined by (\ref{e:et}) and averaged in-plane velocity streamlines defined from (\ref{e:uxz}) for the turbulent spot used as a germ in growth experiments.
The germ has been shifted at the centre of the domain by taking advantage of the periodic boundary conditions. The streamwise direction is horizontal.
Color coding for $e(x,z,t)$ to be used everywhere in the following: from blue for $e\approx 0$ to yellow and next to red for $e\ge0.1$.
\label{f:germ}}
\EF

Figure~\ref{f:et} recapitulates the findings in terms of the distance $\Delta$ to laminar flow as a function of $t$ (in units of the advection time  $h/U$ as used everywhere in the following) for different values of \RE.
\BF
\BC
\includegraphics[width=0.9\textwidth]{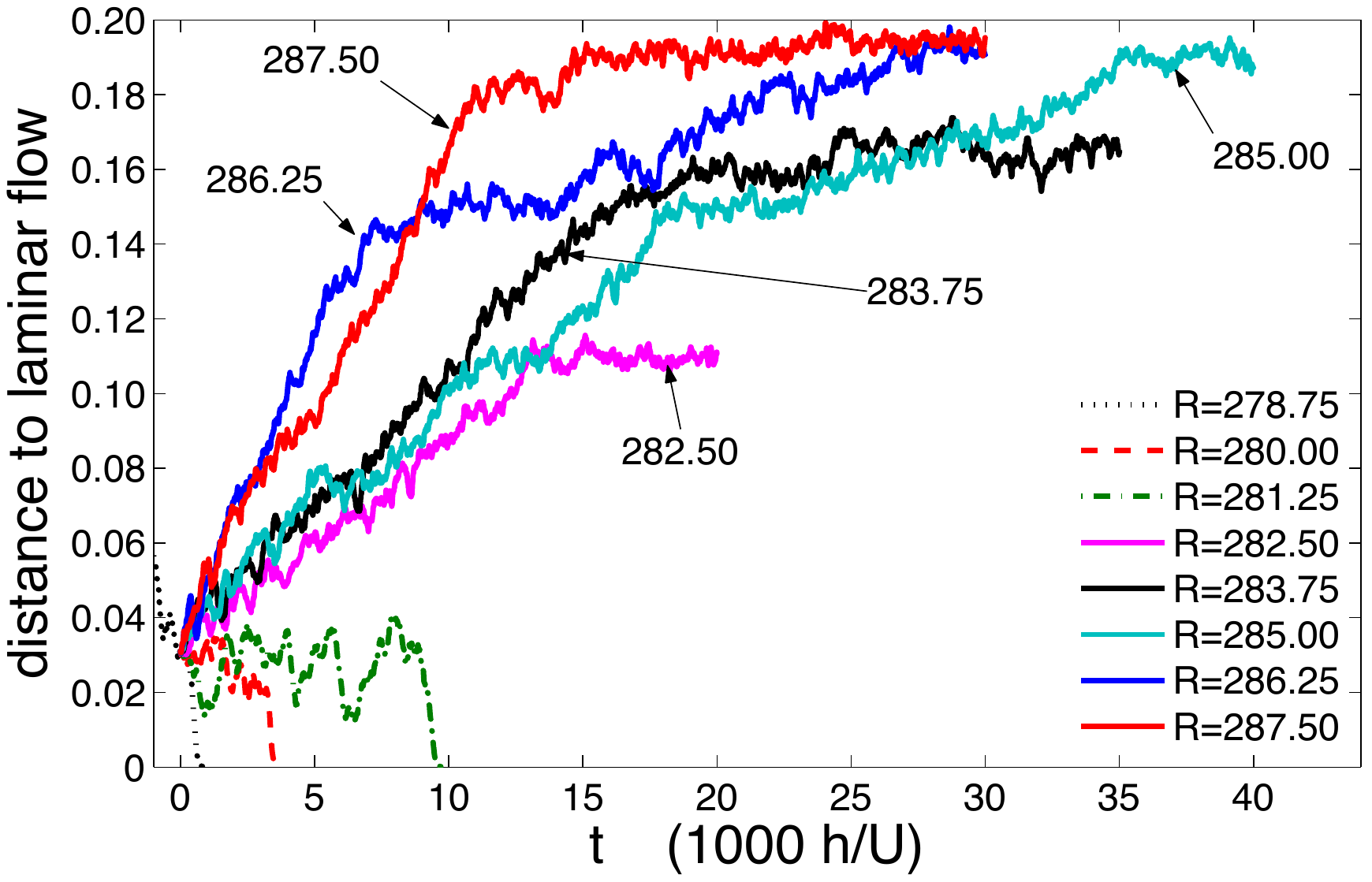}
\EC
\caption{Time series of the distance to laminar flow as a function of time for several Reynolds numbers from the same initial condition.\label{f:et}}
\EF
The first observation is that,  for the values of \RE\ considered, all above $\RG\approx 275$, mature spots in fact decay in a quite short time for $\RE< 282.50$. The second observation is that for $\RE\ge282.5$, the system can reach several asymptotic states with different values of the distance to laminar flow. An illustration of states reached when the simulations were stopped is given in Fig.~\ref{f:finst}.
For $\RE\ge285$, statistically steady solutions all correspond to three more or less well-formed bands.
At $\RE=283.75$ a complicated, unsteady, disorganised pattern with two to three bands is obtained.
A single isolated band survives for $\RE=282.50$.
It should be recalled that, from previous experiments (Manneville 2011), three stable continuous bands were systematically obtained upon decreasing \RE\ adiabatically. The slight change in the dimensions of the domain, from (468,272) here to (432,256) there, is likely to explain that bands were less prone to irregular widths in the smaller domain, when compared to what is seen in Fig.~\ref{f:finst} for $\RE\ge285$. The smaller domain is indeed presumably closer to fitting integer multiples of the patterns' optimal wavelengths in the range of $\RE$ considered here. 
A last remark is that, still from the consideration of variations of $\Delta$ with time, growth can take place at different speeds with plateau stages. The rest of the article aims at interpreting these observations in terms of individual, competing or cooperating processes.

\BF
\includegraphics[angle=90,width=0.19\textwidth]{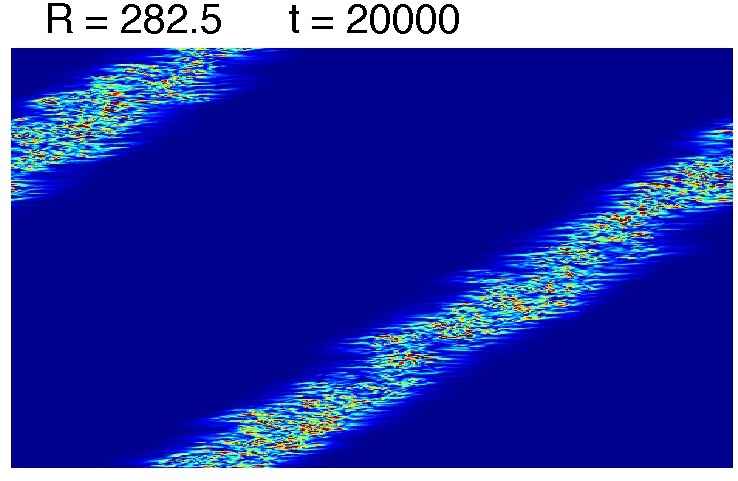}\hfil
\includegraphics[angle=90,width=0.19\textwidth]{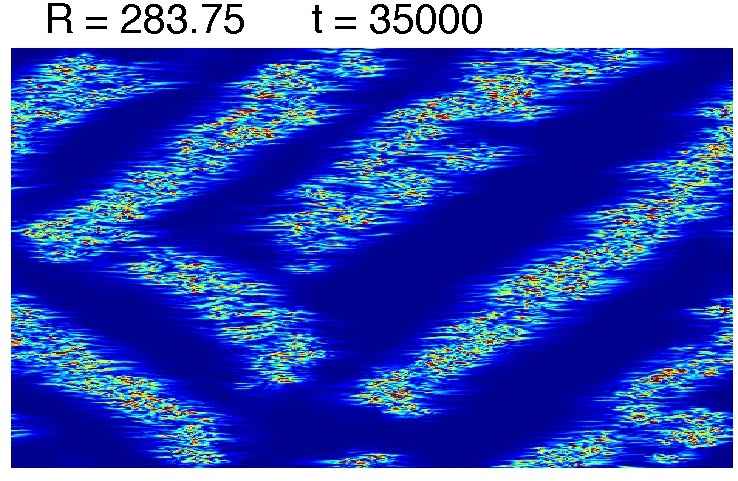}\hfil
\includegraphics[angle=90,width=0.19\textwidth]{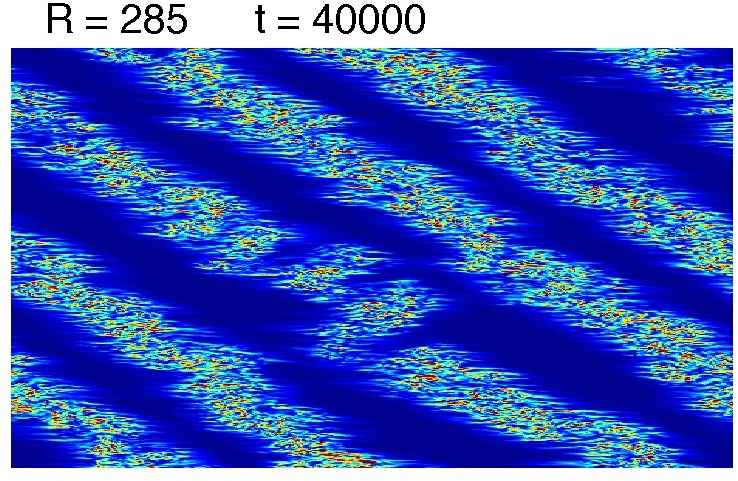}\hfil
\includegraphics[angle=90,width=0.19\textwidth]{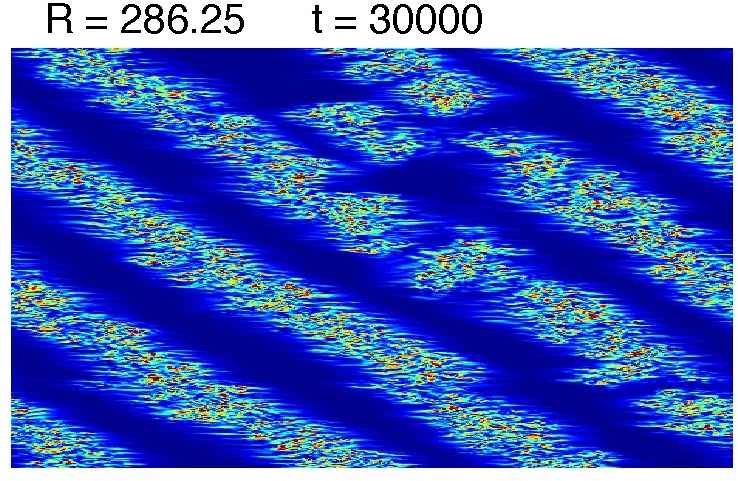}\hfil
\includegraphics[angle=90,width=0.19\textwidth]{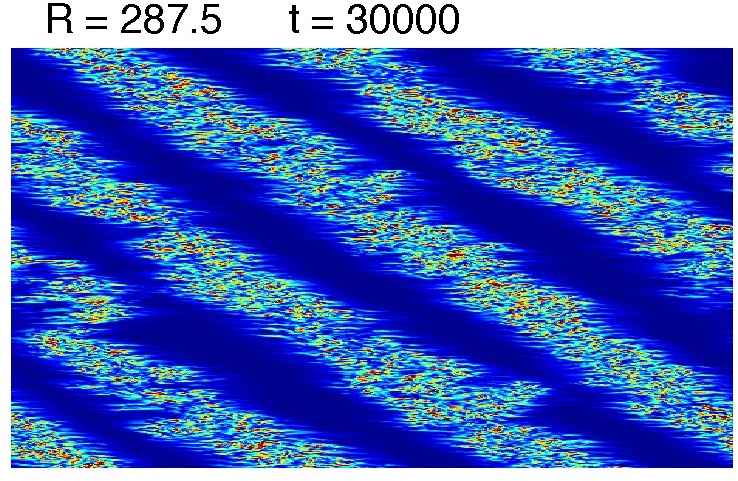}
\caption{States reached when simulations were stopped for, from left to right, $\RE=282.5$ (single band), $\RE=283.75$, $\RE=285$, $\RE=286.25$, and $\RE=287.5$. The streamwise direction is vertical. Same colour coding as in Fig.~\ref{f:germ}.\label{f:finst}}
\EF

\subsection{Growth at $\RE=285$}

This experiment best exemplifies the different processes at work and the different stages observed in wide enough domains with periodic boundary conditions. On general grounds several qualitatively different periods can be distinguished, characterised by the occurrence of specific events. Borrowing the terminology from Duguet {\it et al.} (2011), we call  {\it progress event\/} an episode increasing the turbulent fraction, and {\it retreat event\/} an episode where turbulence breaks down over some region. In contrast with processes observed by Duguet {\it et al.} generally involving a few streaks, our events rather develop at the scale of the width of the turbulent patch considered as illustrated below. The phenomena that we shall encounter will happen repeatedly, throughout the growth process.
\BI
\item Early after the start of the experiment, the system attempts to grow a turbulent branch oriented along the direction symmetrical to that of the initial oblique bar with respect to the streamwise axis. Figure~\ref{f:budding} illustrates this nucleation process that we call {\it budding}. In that case, the bud immediately breaks down and the net result is a longer turbulent segment along the original direction.

\BF
\includegraphics[angle=90,width=0.19\textwidth]{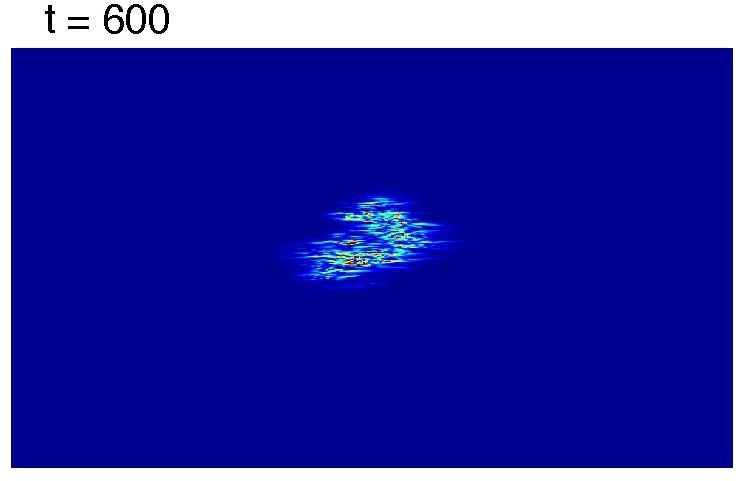}\hfil
\includegraphics[angle=90,width=0.19\textwidth]{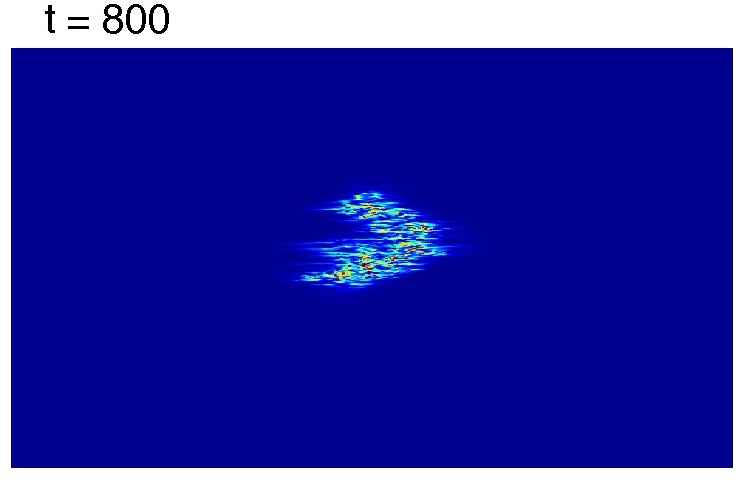}\hfil
\includegraphics[angle=90,width=0.19\textwidth]{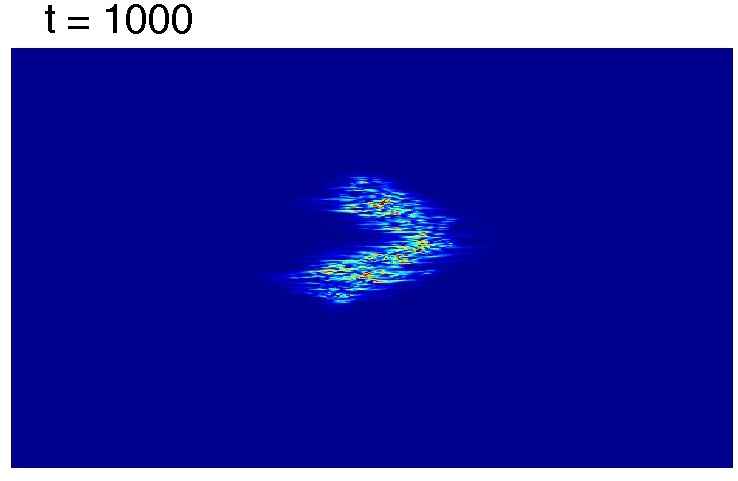}\hfil
\includegraphics[angle=90,width=0.19\textwidth]{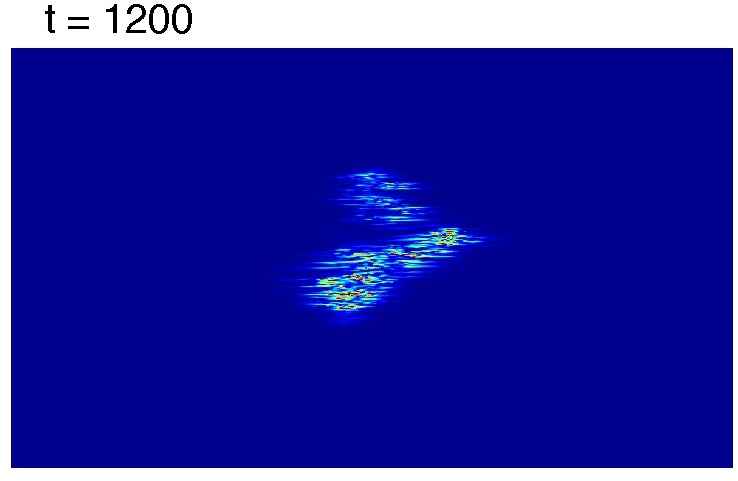}\hfil
\includegraphics[angle=90,width=0.19\textwidth]{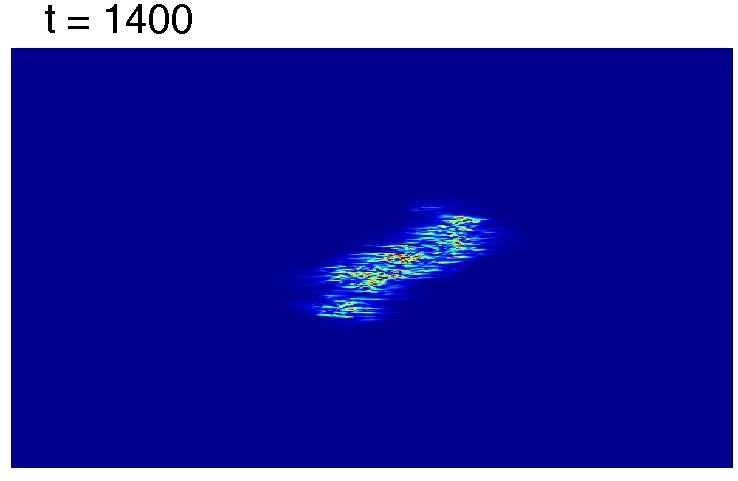}
\caption{Turbulent budding at the start of the experiment for $\RE=285$. Times are indicated to the left of the corresponding images.  The streamwise direction is vertical.\label{f:budding}}
\EF

\item Soon after, another growth process takes place, reminiscent of {\it spot splitting\/} in plane Poiseuille flow (Carlson {\it et al.} 1982) or {\it puff splitting\/} in pipe flow (Avila {\it et al.} 2011), namely the development of a small turbulent patch parallel to the original turbulent bar and separated from it by a quasi-laminar zone. This process is depicted in Fig.~\ref{f:splitting}. Here again, the newly created turbulent zone breaks down immediately while the length of the patch has increased.

\BF
\includegraphics[angle=90,width=0.19\textwidth]{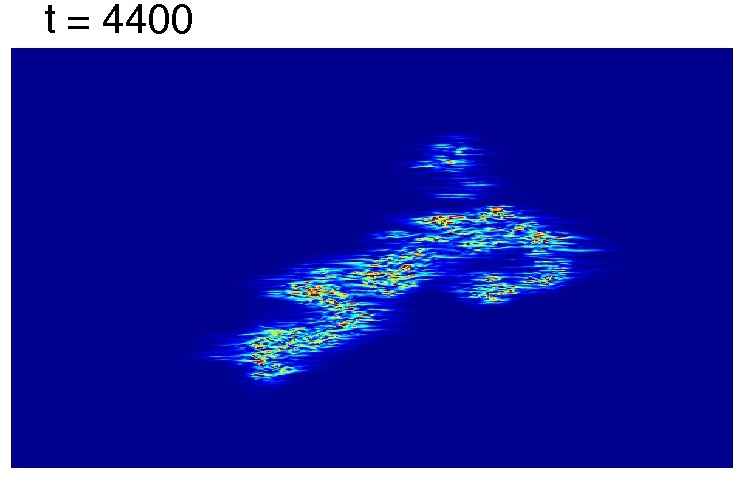}\hfil
\includegraphics[angle=90,width=0.19\textwidth]{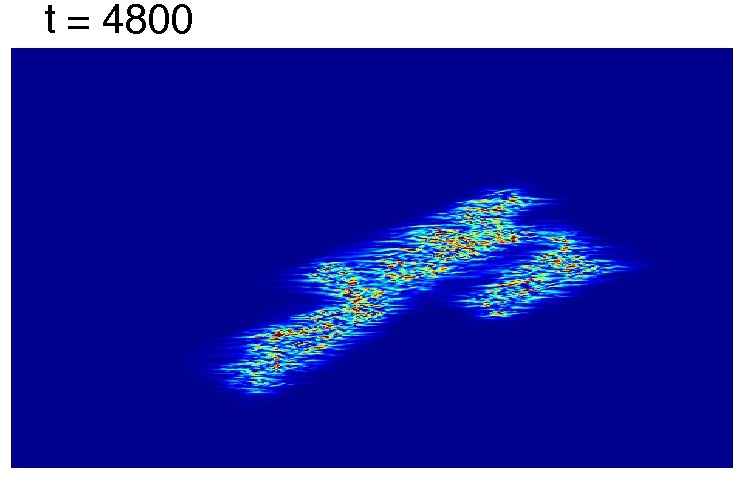}\hfil
\includegraphics[angle=90,width=0.19\textwidth]{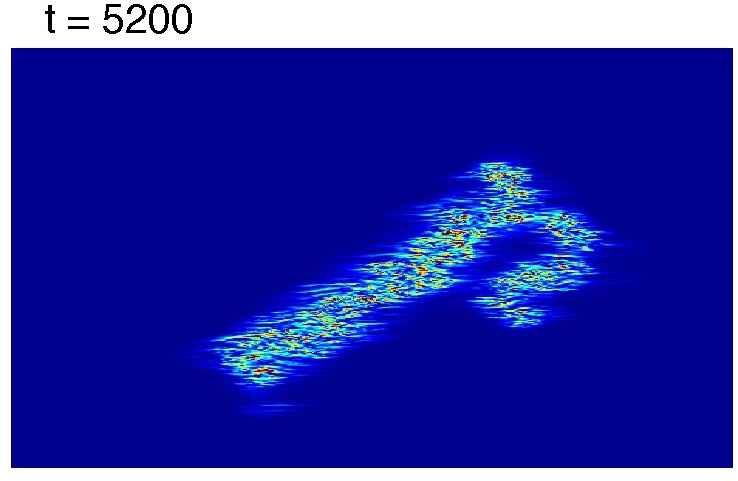}\hfil
\includegraphics[angle=90,width=0.19\textwidth]{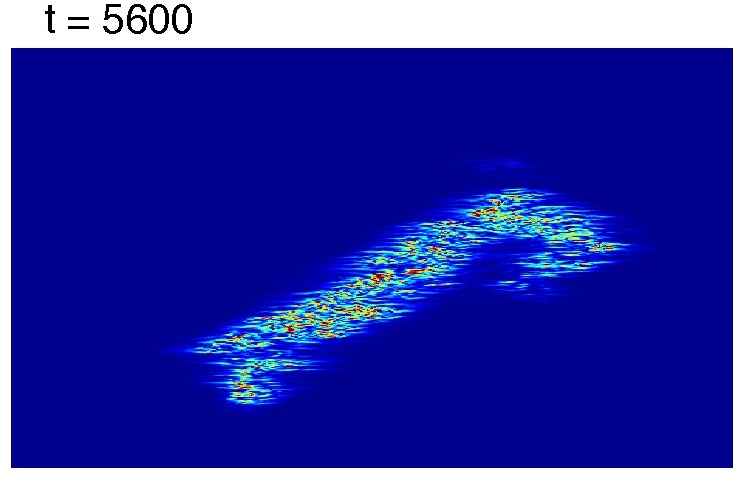}\hfil
\includegraphics[angle=90,width=0.19\textwidth]{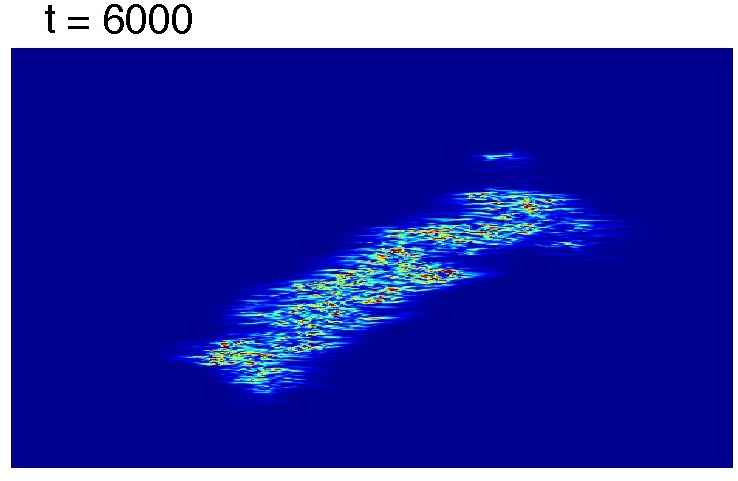}
\caption{Formation of a small turbulent bar next to the germ, here called {\it splitting}. $\RE=285$.   The streamwise direction is vertical.\label{f:splitting}}
\EF

\item At about $t=8\times 10^3$, a budding event starts but the newly created branch no longer decays and a chevron pattern is obtained, see Fig.~\ref{f:laby1} (left and centre-left). This chevron can still be considered as an isolated object as suggested by the representation of the intensity of the mean in-plane flow (central image).
\BF
\includegraphics[angle=90,width=0.19\textwidth]{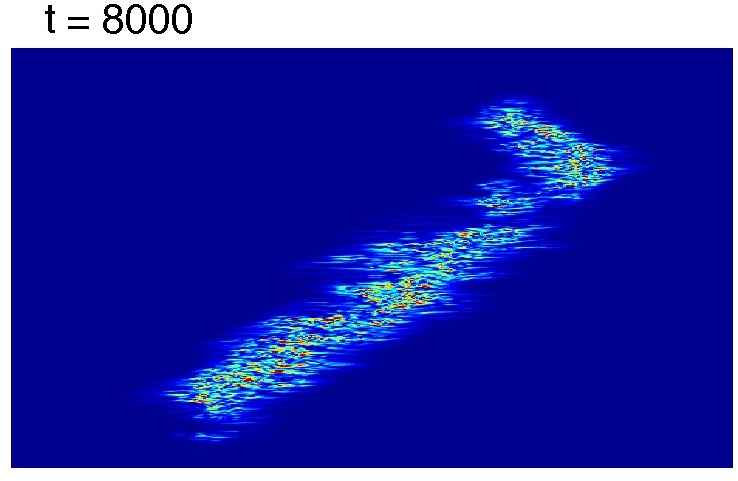}\hfil
\includegraphics[angle=90,width=0.19\textwidth]{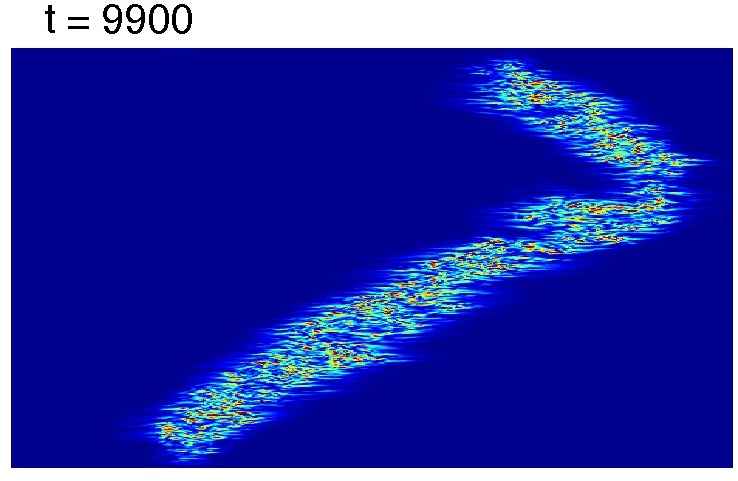}\hfil
\includegraphics[angle=90,width=0.19\textwidth]{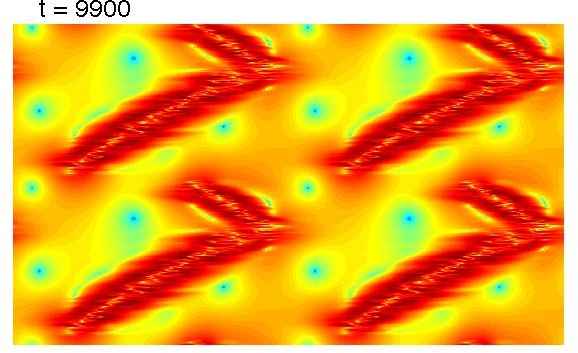}\hfil
\includegraphics[angle=90,width=0.19\textwidth]{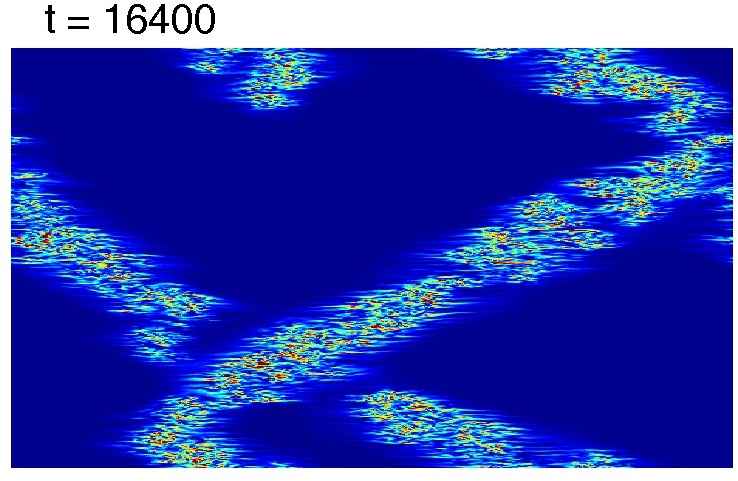}\hfil
\includegraphics[angle=90,width=0.19\textwidth]{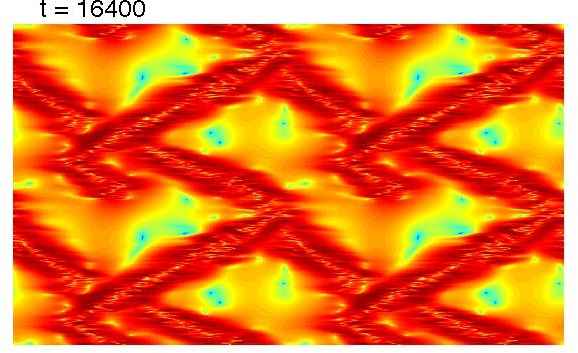}
\caption{Formation of the chevon pattern and next the diamond. The central and right-most images display $2\times2$ tilings of the local intensity of the mean in-plane flow corresponding to the images to their immediate left (decimal logarithms, color scale from blue  ($-10$) to yellow ($-5)$ to red ($-1.7=\log_{10}0.02$). $\RE=285$.   The streamwise direction is vertical.\label{f:laby1}}
\EF

\item Each arm of the chevon next grows along its own direction, until they form a nearly closed diamond-shaped region. The flow inside the diamonds has a complicated pattern but remains extremely weak. Budding of a turbulent segment at the other end of the primary branch at $t\approx1.56\times 10^4$ helps to close the diamond as in the images at $t=1.64\times 10^4$  (Fig.~\ref{f:laby1}, right).
Wide quasi-laminar domains are present at the centre of the diamond and the very same two processes, budding and splitting, operate to fill its interior with banded turbulence.
Compound events frequently occur, i.e. the nucleation of bar-like patches aligned opposite to the local main growth direction and separated from their parents by a small laminar gap, the traces of which are easily identified in the two left-most images in Fig.~\ref{f:laby2}.
The filling is slow but statistically monotonic.
It resembles a trial-and-error process with progress events overcoming retreat events when turbulence collapses over sizeable parts of the newly created turbulent patches.
These processes eventually select a pattern oriented along the direction opposite to that of the initial growth, with the dominant orientation mostly leaning to the left with respect to the streamwise direction (two left-most images  at $t=1.75\times 10^4$, and $t=2.25\times 10^4$), next leaning to the right ($t=2.75\times 10^4$ and $t=3.25\times 10^4$), before eliminating defects to yield a perfect banded pattern, here illustrated at $t=3.75\times 10^4$.

\BF
\includegraphics[angle=90,width=0.19\textwidth]{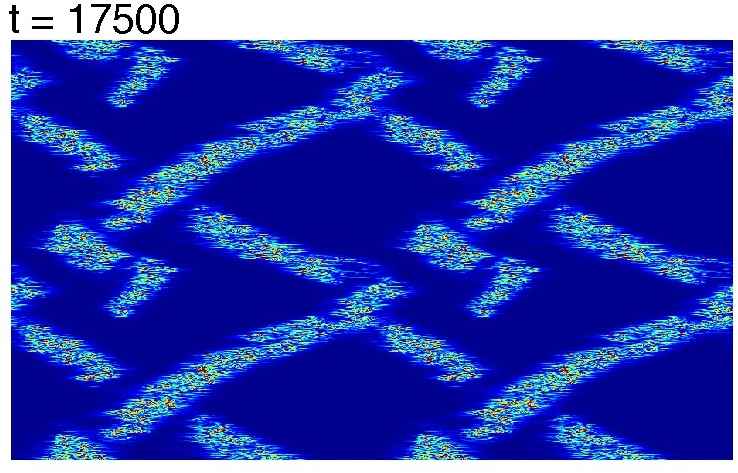}\hfil
\includegraphics[angle=90,width=0.19\textwidth]{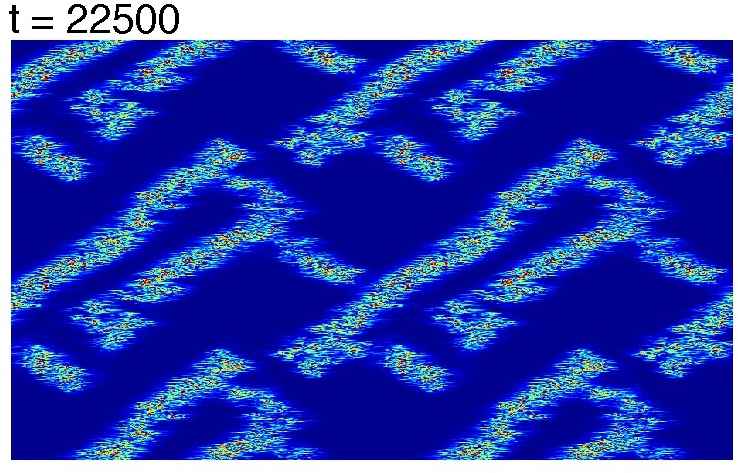}\hfil
\includegraphics[angle=90,width=0.19\textwidth]{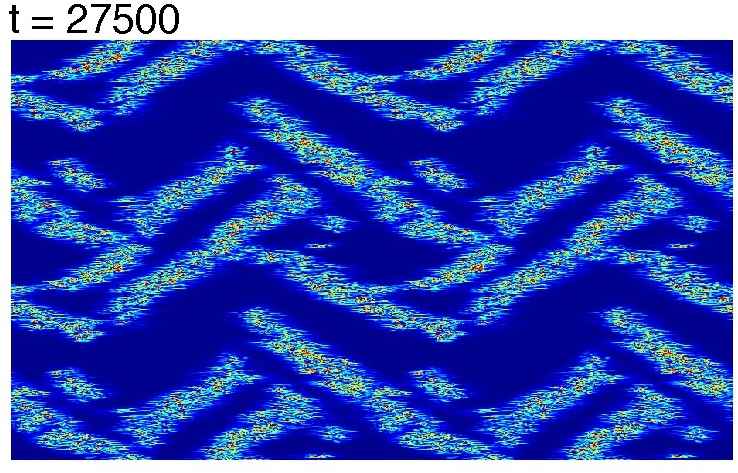}\hfil
\includegraphics[angle=90,width=0.19\textwidth]{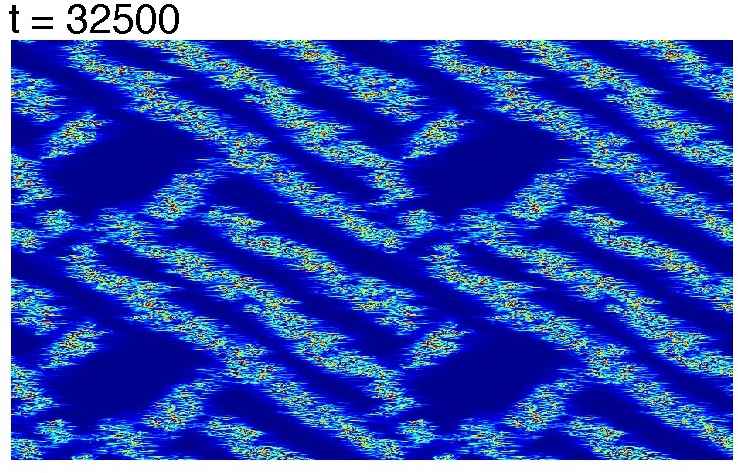}\hfil
\includegraphics[angle=90,width=0.19\textwidth]{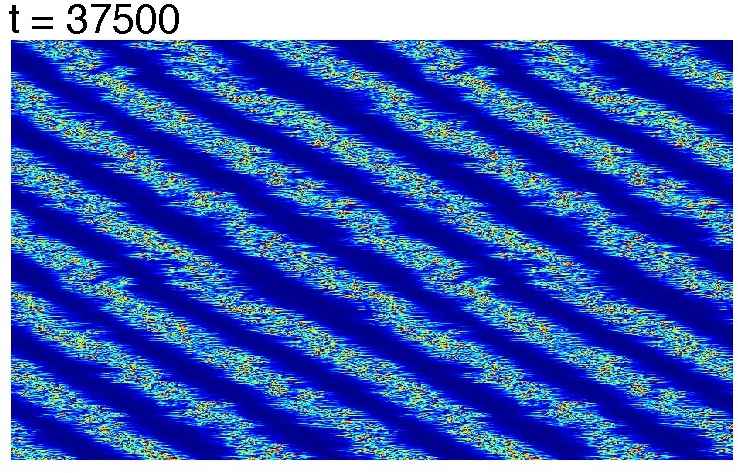}
\caption{Late stage of growth for $\RE=285$ from $t=1.75 \times 10^4$ to $t=3.75\times 10^4$. All images are $2\times2$ tilings of the solution.   The streamwise direction is vertical.\label{f:laby2}}
\EF

\item So, for $t\gtrsim 3.55\times 10^4$ a nearly steady state with three well-formed bands prevails (six when taking the $2\times2$ tiling of Fig.~\ref{f:laby2} into account), however with a trace of instability of the width of the turbulent bands.
This instability evolves into small transversely oriented band fragments that temporarily and locally perturb a given band, cf. image at $t=4\times 10^4$ in Fig.~\ref{f:finst} (centre), which seems to be the way the system finds to solve the mismatch between  the size of the domain and its optimal wavelength at the corresponding value of \RE.

\EI

\subsection{Growth for $\RE\ge282.50$}

Let us first consider the cases with $\RE>285$, namely $\RE=286.25$ and $\RE=287.50$. The faster increase of the distance to laminar flow observed in Fig.~\ref{f:et} can be attributed to a faster spanwise growth, itself due to a smaller probability of turbulent bud decay, so that a connected chevron pattern is observed much earlier.
For $\RE=286.25$, the system gets stuck from $t\approx7.5\times 10^3$ to $1.5\times 10^4$ in a configuration similar to what is observed for $\RE=285$ at $t=1.75\times 10^4$ (Fig.~\ref{f:laby2}, left image) with a wide laminar domain that the system finds difficult to fill.  This correspond to the plateau seen on the corresponding time series of the distance to laminar flow in Fig.~\ref{f:et}. The slow growth that follows the plateau is quite similar to the behaviour recorded for $\RE=385$, with the same processes involved.
In contrast, for $\RE=287.5$, the growth is much more regular, with just a saturation when the three-band configuration is reached.

In both cases, the late growth stage to the defect-free pattern configuration is much slower than the reorganisation stage observed at similar $\RE$ during the experiment starting from featureless turbulence (Manneville 2011) where laminar troughs were progressively created in a uniformly turbulent system with adapted wavelength from the start and no anomalously wide laminar regions.
In contrast, here pattern formation crucially involves turbulent patches in the form of bars growing at their extremities along their main axis, which generates wide laminar domains to be reduced by large deviations processes only (budding/splitting).

We next turn to the cases with $\RE<285$, namely $\RE=283.75$ and $\RE=282.50$. The early and medium growth stages closely correspond to what is observed for $R=285$ in terms of time-series of the distance to laminar flow (Fig.~\ref{f:et}).
However slight frequency differences in the processes at work drive the system to unsaturated patterns, i.e. with turbulent fractions smaller than what could be expected from earlier studies (Manneville 2011). This is due to a larger decay rate for turbulent patches issued from budding and splitting, which mainly limits the possibility of branching while the growth speed along the main growth direction remains approximately constant. This decay rate increase as $\RE$ is decreased which has important consequences on the pattern obtained in the long-time limit:
\BI
\item For $\RE=283.75$  (Fig.~\ref{f:r28375}), the chevron pattern in some sense nearly degenerates with a dominant nearly connected diagonal band and short lateral branches ($t=1.2\times 10^4$). 
Lateral branching occurs repeatedly and a complicated pattern results, following the same trends as when $\RE$ is larger, though the system experiences more difficulty in filling all the laminar patches.
As a result, the turbulent fraction saturates at a smaller value than when perfect or nearly perfect patterns are reached for $\RE\ge285$.
Furthermore, the defective pattern at $t=3.5 \times 10^4$ has returned close to its state at $t=2\times 10^4$, while other configurations with somewhat different topology but roughly the same turbulent fraction, e.g. the state at $t=2.75\times 10^4$, have been visited in the meantime.
This does not prove that the time-asymptotic regime is unsteady due to frustration hampering the regularisation of the pattern, but suggests that the transient towards the permanent regime, if steady, is extremely long due to the long time-scales involved in the large deviation processes at work. 

 \BF
\includegraphics[angle=90,width=0.19\textwidth]{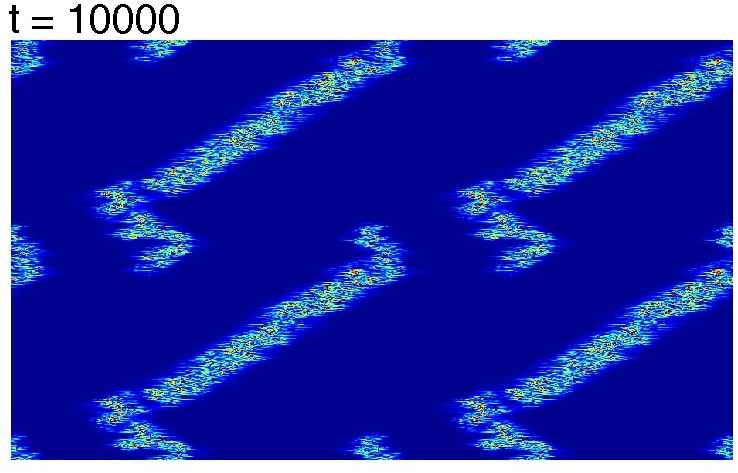}\hfil
\includegraphics[angle=90,width=0.19\textwidth]{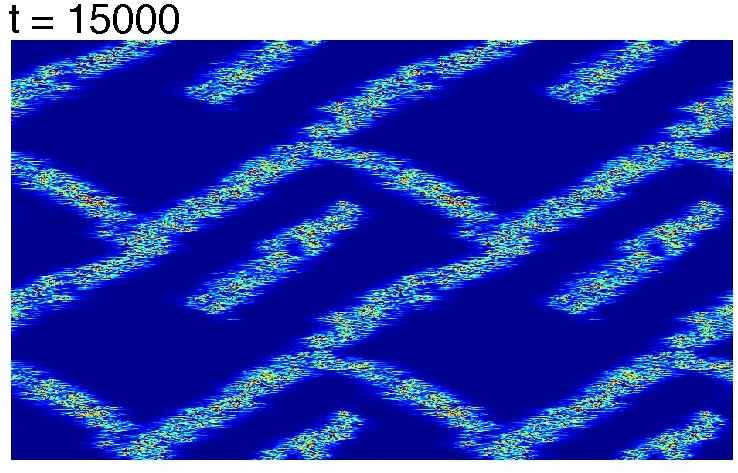}\hfil
\includegraphics[angle=90,width=0.19\textwidth]{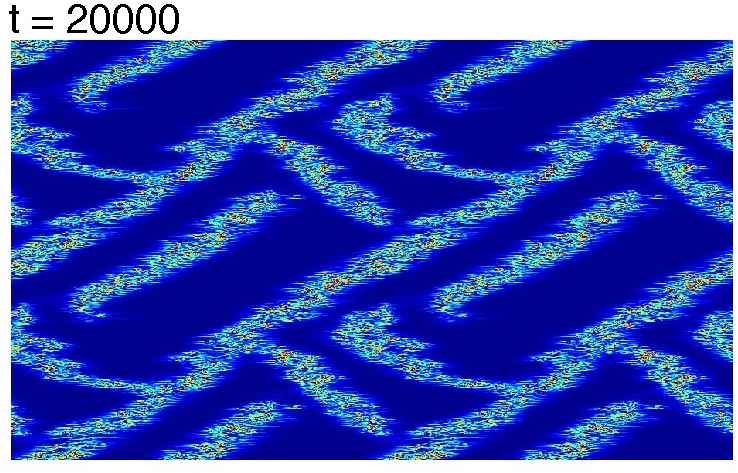}\hfil
\includegraphics[angle=90,width=0.19\textwidth]{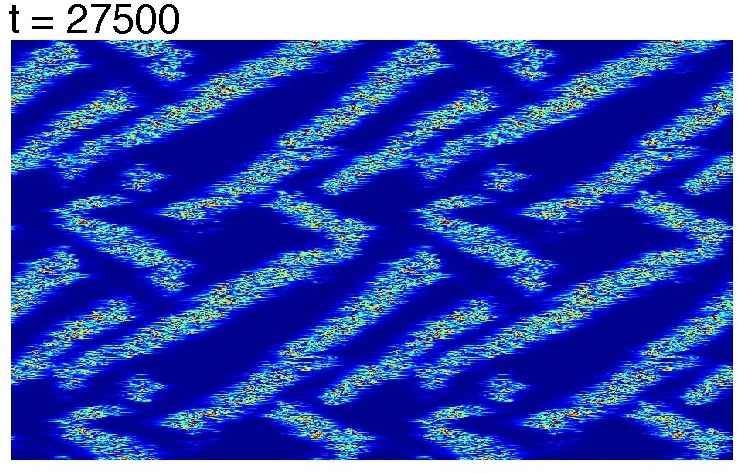}\hfil
\includegraphics[angle=90,width=0.19\textwidth]{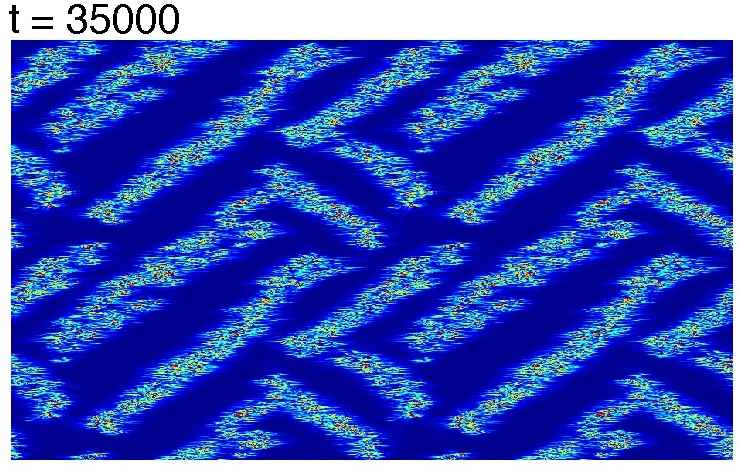}
\caption{Late stage of growth at $\RE=283.75$ from $t=10^4$ to  $t=3.5\times 10^4$  ($2\times2$ tilings). At $t=3.5\times 10^4$ the system has returned close to its state at $t=2\times 10^4$.  The streamwise direction is vertical.\label{f:r28375}}
\EF

\item For $\RE=282.5$, whereas budding and splitting still take place at roughly the same rate, the decay of newly created turbulent  patches is  large enough that the chevron cannot develop, since the lateral branch is destroyed before having a chance to produce a nontrivial pattern.
Since the turbulent segment still continues to grow, a single band is obtained in the long-time limit.
Though specific to the in-plane periodic boundary conditions, the interesting phenomenon here relates to how the large-scale quasi-laminar flow around the turbulent region reconnects (Fig.~\ref{f:r28250}), as such reconnection processes also take place in the other cases. At $t=10^4$ (top-left images) a sizeable gap is open with the $(\bar u_x,\bar u_z)$ field displaying a saddle point. At $t=1.28\times 10^4$ (top-right images), the gap is being filled with turbulence.
The large-scale flow is most intense at the turbulent laminar interface and clearly affects the budding process that took place a little earlier ($t=1.17\times 10^4$). It is also seen to play a role in the elimination of the irregularities of the width of the turbulent band just after the reconnection, leaving us with a nice continuous band here featured at $t=2\times 10^4$ (Fig.~\ref{f:r28250}, bottom images). Barkley \& Tuckerman (2005, 2011) found such isolated band states using their skewed-streamwise confined domain close to the global stability threshold. The present observation shows that such states are possible stable configurations also in weakly constrained periodic domains provided that the aspect-ratios are compatible with the preferred band angle at the relevant Reynolds number. 
\EI

 \BF
 \BC
\includegraphics[angle=90,width=0.19\textwidth]{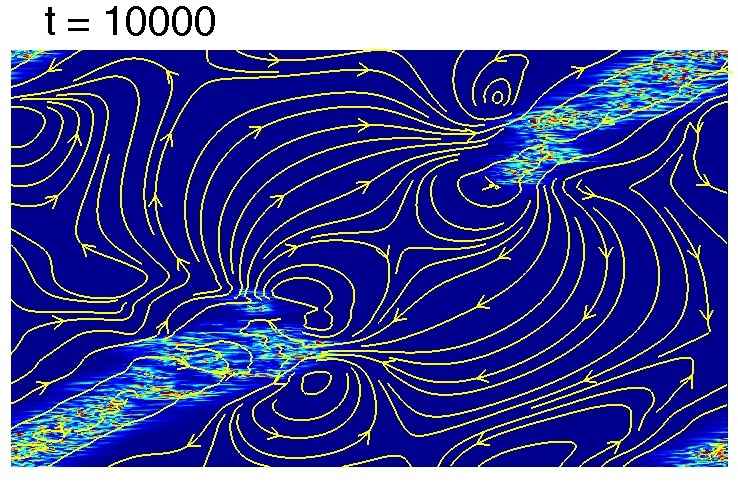}\hspace{0.5em}
\includegraphics[angle=90,width=0.19\textwidth]{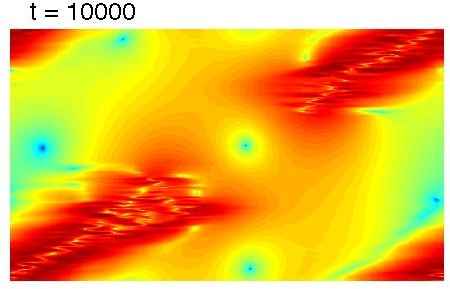}\hspace{3.5em}
\includegraphics[angle=90,width=0.19\textwidth]{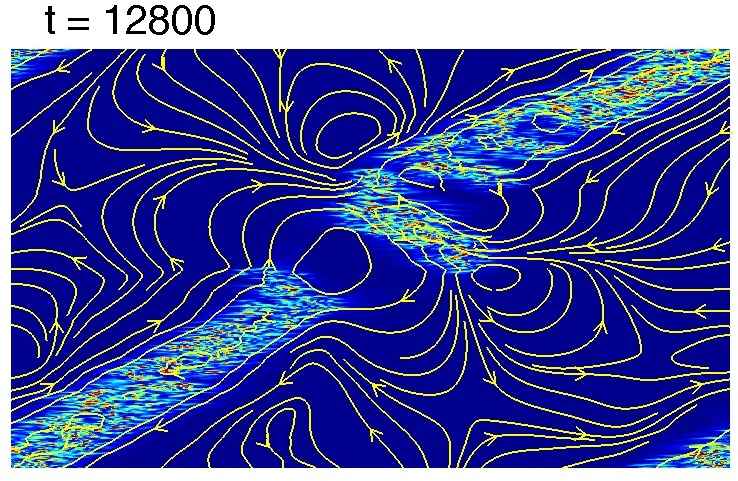}\hspace{0.5em}
\includegraphics[angle=90,width=0.19\textwidth]{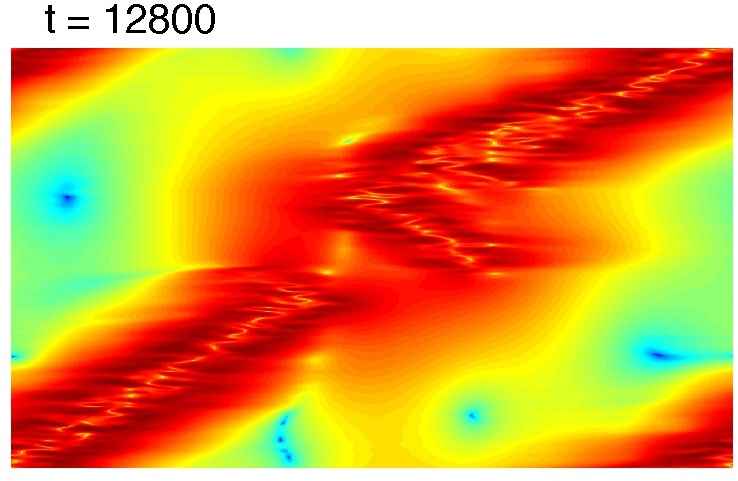}\\[2ex]
\includegraphics[angle=90,width=0.19\textwidth]{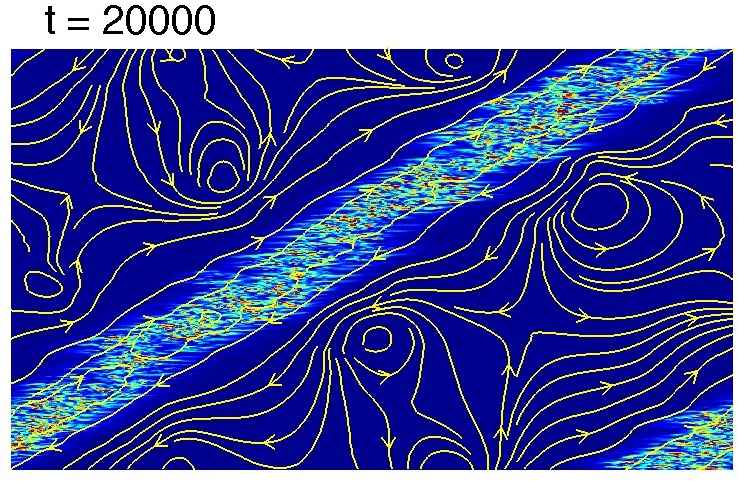}\hspace{0.5em}
\includegraphics[angle=90,width=0.19\textwidth]{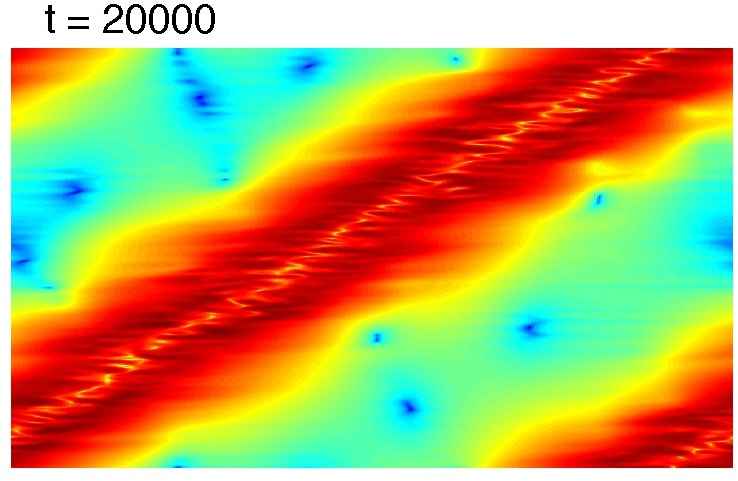}
\EC
\caption{Late stage of growth at $\RE=282.50$. In each group, the local distance to laminar flow and the streamlines of the $(\bar u_x,\bar u_z)$ field  are depicted in the left image and the intensity of the $(\bar u_x,\bar u_z)$ field in the right image (decimal logarithms, colour scale as in Fig.~\ref{f:laby1}). Top-left: $t=10^4$, the two tips are separated by a laminar gap. Top-right: $t=1.28\times 10^4$, reconnection. Bottom: $t=2\times 10^4$,  continuous band.   The streamwise direction is vertical. \label{f:r28250}}
\EF

\subsection{Decay for $\RE\le281.25$}
No sustained pattern has been obtained for $\RE<282.50$ when starting from the germ displayed in Fig.~\ref{f:germ}.
At $\RE=281.25$, laminar flow is recovered at the end of the interesting long transient illustrated in Fig.~\ref{f:r28125}. 
During this transient, the turbulent fraction falls down to quite small values at $t\approx800$, $4300$ ($\circ$), or $6500$ ($\circ$) but the systems recovers with turbulent patches at least as large as the initial germ at $t\approx1800$, $2500$, $4000$, $5700$ ($\circ$), or $8000$ ($\circ$), experiencing a rapid collapse after $t\approx9200$ ($\circ$).
Similar behaviour is observed for $\RE=280$ and $278.75$ but transients are much shorter (see Fig.~\ref{f:et}) due to an increase of the collapse probability when $\RE$ is decreased. This is in line with Bottin's study of triggered spots (Bottin 1998). Any statistical analysis is however still out of reach since it would be necessary to vary the size, the shape and the intensity of the initial germ, which is unfeasible, even with our under-resolved numerics.

\BF
\BC
\includegraphics[width=0.6\textwidth]{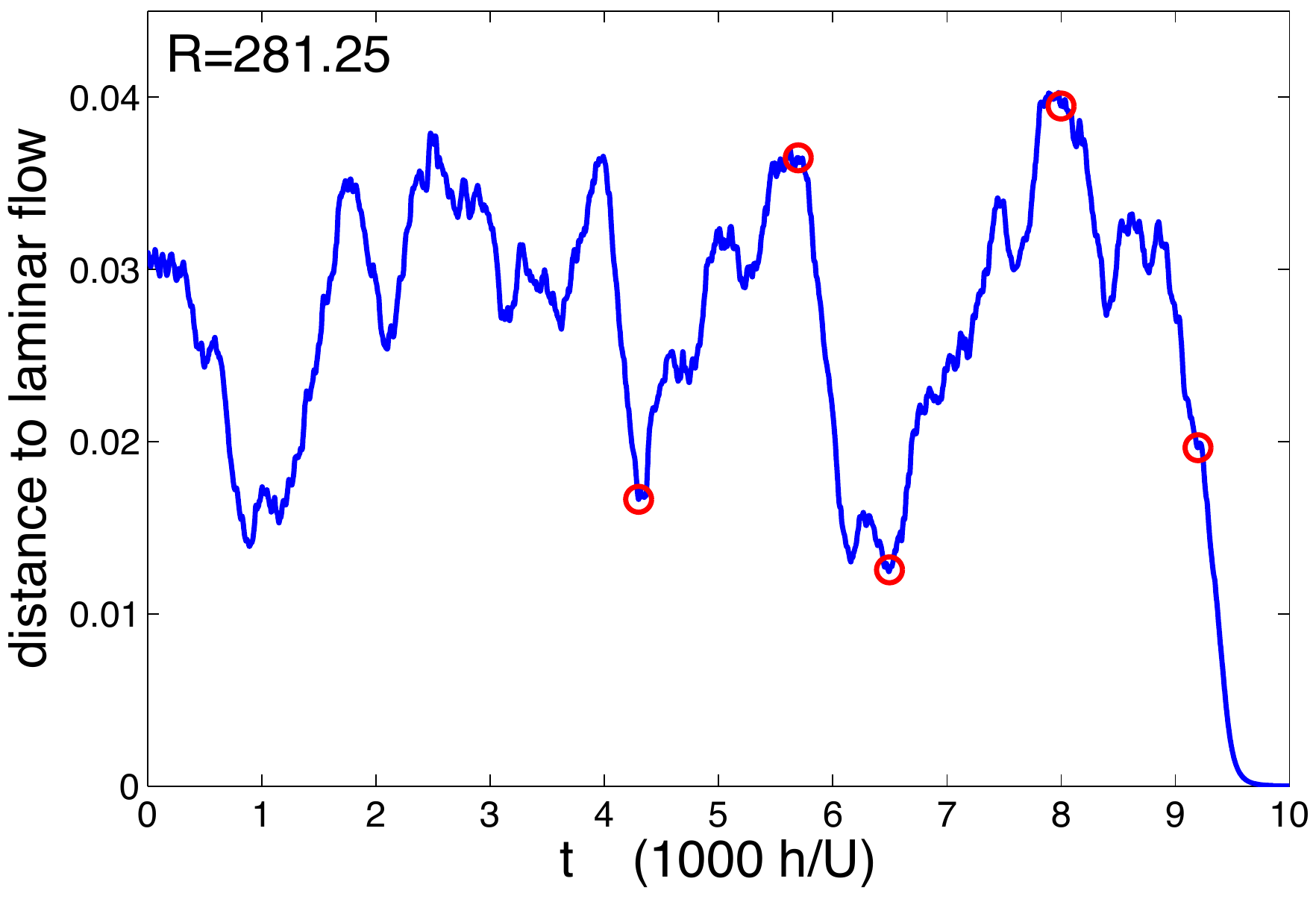}
\EC
\includegraphics[angle=90,width=0.19\textwidth]{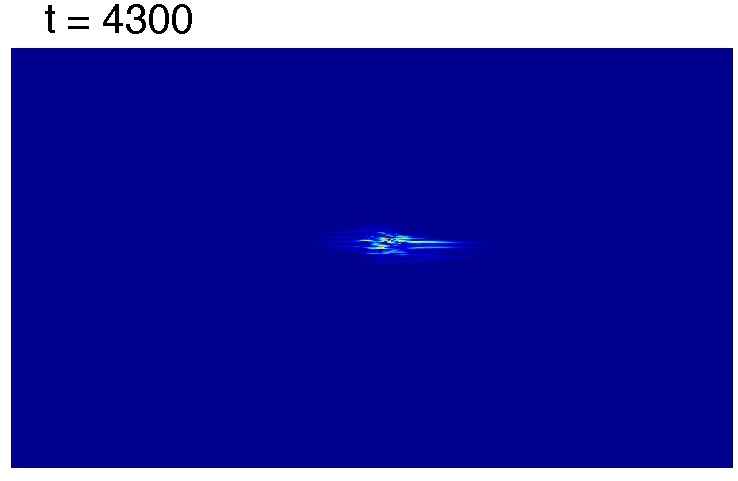}\hfil
\includegraphics[angle=90,width=0.19\textwidth]{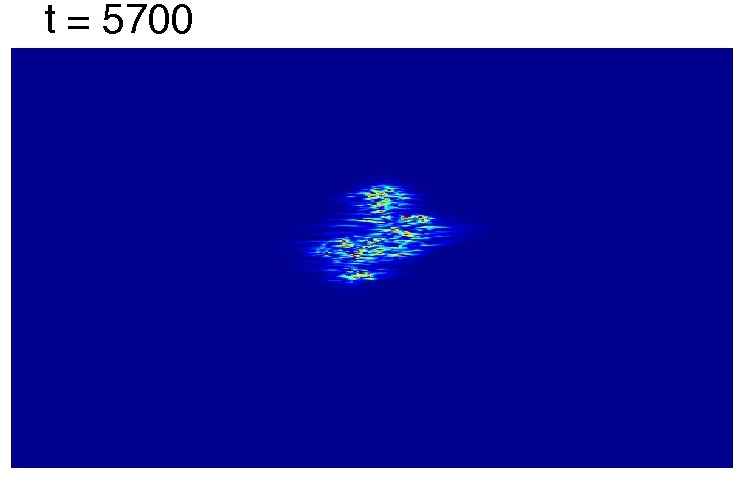}\hfil
\includegraphics[angle=90,width=0.19\textwidth]{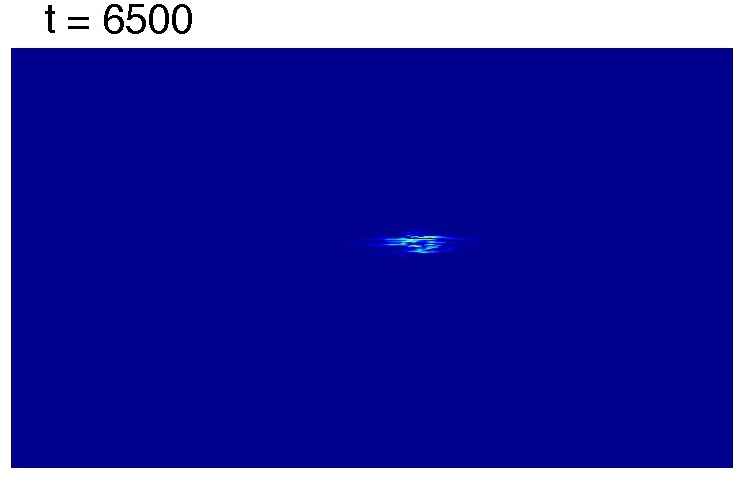}\hfil
\includegraphics[angle=90,width=0.19\textwidth]{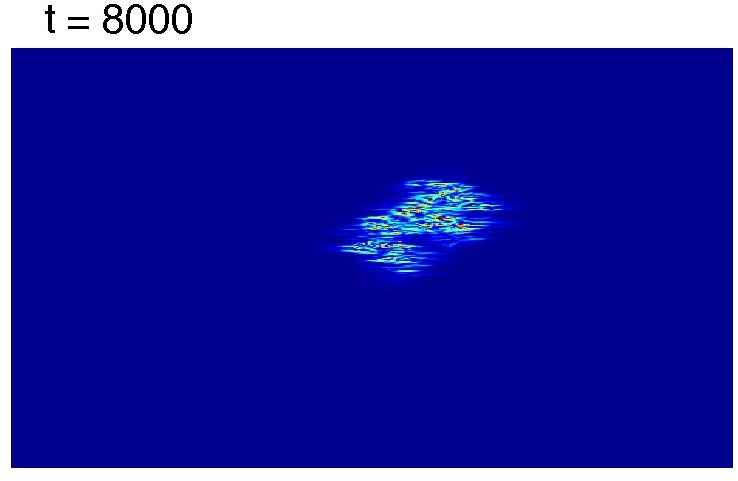}\hfil
\includegraphics[angle=90,width=0.19\textwidth]{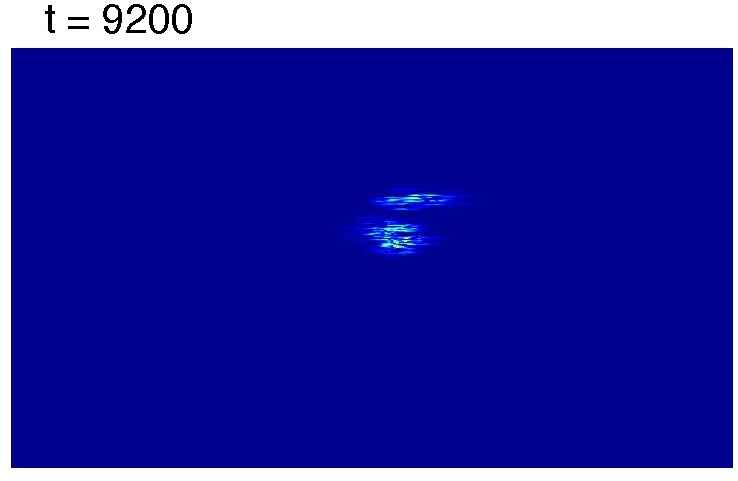}
\caption{Decay of the germ for $\RE=281.25$. Top: Time series of the distance $\Delta$ to laminar flow. Bottom: Snapshots of the solution at the instants identified by open circles ($\circ$) in the graph.   The streamwise direction is vertical.\label{f:r28125}}
\EF

\section{Discussion\label{sec3}}

In Section~\ref{sec2} above, we have presented the phenomenology of turbulent pattern growth in plane Couette flow.
Direct numerical simulations at reduced resolution were performed in domains wide enough to accommodate several bands. 
All the qualitative characteristics of the laminar-turbulent transition were preserved, in particular the self-sustaining process (Waleffe 1997) at the scale of the width of a pair of streaks (minimal flow unit or MFU, i.e. {\it microscopic\/}), transient chaos in the transitional range (Eckhardt {\it et al.} 2008), and the {\it mesoscopic\/} organisation of turbulence in bands (Prigent {\it et al.} 2003), at just the price of a slight but systematic discrepancy in the quantitative predictions, such as the observed 15\% downward shift of the $[\RG,\RT]$ interval.
Accordingly, we focused on the identification of qualitative processes rather than on quantitative statistical estimates.
Our observations give unambiguous support to seminal ideas put forward by Pomeau (1986) and further developed by him (1998) about the transition to turbulence in globally subcritical systems.

In Pomeau's views, state coexistence implied by subcriticality has crucial consequences when dealing with the laminar-turbulent transition in extended geometry.
In this case, lateral boundary conditions at large distances are sufficiently weak to allow an understanding of the flow in terms of some spatiotemporal dynamics defined on a lattice of coupled local sites evolving among several possible states.
Fronts expected to separate homogeneous domains of sites in different states were conjectured to fluctuate when one of these states was chaotic while the other was regular.
Accordingly, {\it directed percolation\/} --~a stochastic contamination process studied in statistical physics (Hinrichsen 2000)~-- was proposed as an appropriate framework to discuss the phenomenon. In the transitional range of wall-bounded flows, laminar flow is locally stable and would quite naturally be the {\it absorbing\/} state, which can evolve towards nontrivial dynamics only by {\it contamination\/} from neighbours; in turn, locally turbulent flow would  of course play the role of the {\it active\/} state.
Introducing transiently chaotic local dynamical sub-systems coupled in a chain, he then proposed an implementation of this statistical physics concept in terms of {\it spatiotemporal intermittency\/}, see Chat\'e \& Manneville (1988) for a concrete application.

A second ingredient of Pomeau's approach is rooted in an analogy between the subcritical transition to turbulence in extended systems and {\it first-order phase transitions\/} in thermodynamics (such as the solid--liquid transition), and more particularly with the associated kinetic aspects involving the {\it nucleation\/} of one phase inside the other (Pomeau 1998):
The new phase develops around {\it germs\/} and the transition takes place when local fluctuations larger than some {\it critical germ\/}  are present, the new phase then invading the system.
This framework was tested by Bottin \& Chat\'e (1998) within the framework of spatiotemporal intermittency in the case of a discontinuous transition, and shown to be relevant  (Manneville 2009) in a model of plane Couette flow partially accounting for its subcritical transition (Lagha \& Manneville 2007a).
In both cases the non-trivial role of streamwise anisotropy and turbulent band organisation were however neglected, which is no longer the case here.

Pomeau's educated guesses are indeed well illustrated by our experiments that reproduce the transitional range more realistically than previous modelling attempts (Bottin \& Chat\'e 1998; Lagha \& Manneville 2007a; Manneville 2009).
During decay experiments reported in (Manneville 2011), two basic processes were pointed out, one {\it microscopic\/} (MFU scale): the withdrawal of band fragments at their extremities, and the other {\it mesoscopic\/} (band width scale): the opening of a sufficiently wide laminar gap inside a long enough band segment.
In growth experiments studied here, two complementary basic processes with similar characteristics are present, one microscopic: the growth of band fragments at their tip and along the direction of their length (at the statistical level), and one mesoscopic, in two forms: {\it splitting\/}, nucleation of a turbulent segment locally parallel to the band, but at a distance from it, and {\it budding\/}, transversally oriented turbulent offspring at a band segment extremity.

Probabilities attached to local processes -- collapse of turbulent streaks and contamination of laminar flow by nearby turbulence -- control a phenomenon akin to directed percolation, either biased towards decay for $\RE\lesssim275$ as seen previously (Manneville 2011), or towards growth as seen here for $\RE\gtrsim282.5$.
A threshold   $\RE_{\rm g}'$, similar to that of directed percolation, would then be obtained by optimisation as a lower bound on \RE\ for local growth from a germ with variable size, shape, and turbulent intensity, which at any rate is out of reach with  present numerical capabilities.
One would therefore have $\RE_{\rm g}'\lesssim 282.5$ but $\ge 275$ since decay is effective at \RG.
However the transition need not be strictly analogous to directed percolation, i.e. continuous, second-order, with critical exponents in the same universality class (Hinrichsen 2000),
as there are examples of similarly defined non-equilibrium processes experiencing discontinuous transitions even in one dimension (Dickman \& Tom\'e, 1991).%
\footnote{In contrast, Barkley (2011) indicates that for transitional pipe flow, which effectively behaves as a one-dimensional system,  the critical properties of one-dimensional directed percolation might apply.}

Growth and decay due to large deviations (splitting, budding, combination of both, and turbulence collapse at the scale of the band width) are indeed conspicuous and the overall dynamics can be understood from a detailed examination of properties of individual events, e.g. the occurrence of spanwise growth, the irreversible invasion of the domain by a labyrinth of band segments with two symmetrical orientations, the final regularisation of the pattern by elimination of defects of all sorts, or the fact that, when the decay rate of newly created turbulent patches is too large, the system does not enters the regime of labyrinthine growth but builds up a single band.

The little-constrained quasi-two-dimensional geometry of our experiments offers many more possibilities than the more constrained quasi-one-dimensional geometry of Duguet {\it et al.} (2011) for plane Couette flow [or pipe flow considered by Moxey \& Barkley (2011) or Avila {\it el al.} (2011)].
Such a versatility is accordingly much harder to frame quantitatively at the probabilistic level.
For example, it does not seem possible to define a threshold by just comparing the probability for retreat events and a single type of progress event as done by  Duguet {\it et al.} (2011), in analogy with  decay and splitting rates  considered by Avila {\it et al.} (2011).
The many specific processes discussed above are undoubtedly too small-scale to be reliably estimated at the quantitative level with our simulations at  reduced resolution.
However, this does not reduce the generic character of their relevance.

As a last remark, let us consider decay from a germ, here observed at $\RE\le281.25$.
First, the general behaviour during long transients (see Fig.~\ref{f:r28125}) is extremely similar to findings of Bottin (1998) and we expect that a statistical study would produce exponentially decreasing lifetime distributions as in the laboratory, but even at reduced resolution, this would imply a tremendous amount of work with little return.
It is however interesting to observe that,  for $\RE$ between 281.25 and 282.5, the chosen germ is clearly on the border of the attraction basin of laminar flow. In contrast with {\it edge states\/} that were obtained much beyond \RG\ but at a small distance from laminar flow (Duguet {\it et al.} 2009), the solution followed here appears far enough from laminar flow mostly because its spatial extension varies wildly.
This might reveal the actual role of these edge states because,  at its closest approach to laminar flow, the solution which is followed here has a spatial structure that, apart from being much less symmetric, is tightly localised and looks similar to observed typical edge states, e.g. at $t=4.3\times 10^3$ and $6.5\times 10^3$.
Edge states are therefore unstable structures visited during long transients.
The corresponding intermittent dynamics is then in two stages, ({\it i\/})~deterministic escape from edge states up to extended enough turbulent patches and ({\it ii\/})~stochastic return to edge states according to the general spatiotemporal dynamics governing turbulence in the transitional regime.
This takes place in a very limited range of Reynolds numbers around \RG\ where growth and decay via large deviations have similar probabilities, hence near complete breakdown and recovery.
Final decay (here for $t\ge9.2\times 10^3$) is eventually understood as due to a large deviation missing the edge state and ejecting the system directly to laminar flow.
These trajectories however live in a region of phase space that is separated from the region corresponding to well-formed bands that, for all practical purposes, can be considered as stable down to \RG.
They find their way to sustained turbulence only at larger \RE\ when growth overcomes decay by a sufficient amount.
Incompletely saturated states can be reached owing to frustration imposed by boundary conditions set at large lateral distances, at least as long as \RE\ is not large enough to ensure defect healing.

To conclude, whereas a large body of work has been devoted to systems confined by periodic conditions at a small distance (MFU)  where small scale coherence and temporal behaviour are important, our study relates to the fully spatiotemporal dynamics of systems in extended geometry more relevant to laboratory experiments, and for which the conceptual framework of phase transitions proposed by Pomeau (1986, 1998) appears most adapted.
In particular, it is suggested that large deviations and nucleation processes, inherent in first-order transitions, govern the transition to/from turbulence, while processes akin to directed percolation drive the details of growth/decay processes.

The main objectives of future work should thus be the elucidation of hydrodynamical mechanisms by which the laminar-turbulent interface maintains itself or moves, and, when it moves, continuously by local processes at the scale of a few streaks at the tip of turbulent segments, or on the contrary by jumps at the scale of several streaks, i.e. budding, splitting or collapse. The role of large scale flows generated by Reynolds stresses inside the turbulent patches should also be scrutinised.

 In view of the close correspondence between our results and experimental findings or fully resolved simulations at moderate aspect ratio (band decay, growth from a germ, general band organisation), we are confident that the results presented here are generic and just shadow the physical situation at Reynolds numbers about 15\% larger.
 Confirmation from fully resolved numerics (as well as dedicated laboratory experiments) would however be welcome, in particular to rank the probabilities of the different processes observed, and possibly define appropriate thresholds as in (Duguet {\it et al.} 2011) or (Avila {\it et al.} 2011).
Finally, beyond the specific case of plane Couette flow, others wall-bounded flows such as channel, rotor-stator, and boundary layer flows, would warrant a similar study, in view of their technical importance.

\paragraph{Acknowledgments.} 
This article covers part of the results presented at  {\it Fourth International Symposium ``Bifurcations and Instabilities in Fluid Dynamics''} in Barcelona (18th--21st July, 2011). We would like to thank D~Barkley for his invitation to contribute to the Special Session on {\it Subcritical Instabilities and Coherent Structures in Shear Flows\/}, for discussions, and for the communication of his results prior to publication.
Thanks are also due to J~Rolland for his collaboration and help on computational issues, Y~Duguet, L~S~Tuckerman, and the members of the Saclay Group {\it Instabilit\'es et Turbulence\/} for many stimulating discussions.
\vspace{4ex}

\noindent{\Large\bf References}

\vspace{2ex}

{\leftskip1em
\parindent-1em
\parskip1ex

 Avila K, Moxey D, de Lozar A, Avila M,  Barkley, D and Hof B 2011
The onset of turbulence in pipe flow {\it Science} {\bf 333} 192--196

 Barkley D and Tuckerman L S 2005
Computational study of turbulent laminar patterns in Couette Flow
{\it Phys. Rev. Lett.} {\bf94} 014502

 Barkley D and Tuckerman L S 2007
Mean flow of turbulent laminar pattern in Couette flow
{\it J. Fluid Mech.} {\bf 574} 109--137

 Barkley D 2011
Simplifying the complexity of pipe flow
{\it Phys. Rev. E\/} {\bf 84}  016309

 Berg\'e P, Pomeau Y and Vidal C. 1998 {\it L'espace chaotique\/} (Hermann, Paris) 
 
 Bottin S 1998 {\it Structures Coh\'erentes et Transition vers la Turbulence par Intermittence
Spatio-temporelle dans l'\'Ecoulement de Couette Plan\/}
PhD thesis, Universit\'e Paris-Sud, Orsay,\\
{\tt http://tel.archives-ouvertes.fr/tel-00001111/en/}

 Bottin S and Chat\'e H 1998
Statistical analysis of the transition to turbulence in plane Couette flow\\
{\it Eur. Phys. J. B\/} {\bf 6} 143--155

\newpage

 Bottin S, Daviaud F, Manneville P and Dauchot O 1998
Discontinuous transition to spatiotemporal intermittency in plane Couette flow
{\it Europhys. Lett.} {\bf 43} 171--176.

  Carlson D R, Widnall S E and Peeters M F 1982 
A flow visualization of transition in plane Poiseuille flow
{\it J. Fluid Mech.} {\bf 121} 487--505

  Chat\'e H and Manneville P 1988
Spatiotemporal intermittency in coupled map lattices
{\it Physica~D\/} {\bf 32} 409--422

 Dauchot O and Daviaud F 1995
Finite amplitude perturbation and spots growth mechanism in plane Couette flow
{\it Phys. Fluids} {\bf 7} 335--343

 Dickman R and  Tom\'e T 1991
First-order phase transition in a one-dimensional nonequilibrium model
{\it Phys. Rev. A} {\bf 44} 4833--4838 

 Duguet Y, Schlatter P and Henningson D S 2009
Localized edge states in plane Couette flow
{\it Phys. Fluids\/}~{\bf 21} 111701

 Duguet Y, Schlatter P and Henningson D S 2010
Formation of turbulent patterns near the onset of transition
in plane Couette flow
{\it J. Fluid. Mech\/} {\bf 650}, 119--129

 Duguet Y, Le Ma\^itre and Schlatter P 2011
Stochastic and deterministic motion of a laminar-turbulent front in a spanwisely extended Couette flow
{\it Phys. Rev. E\/} {\bf84} 066315

 Eckhardt B, Faisst H, Schmiegel A \& Schneider T.M. 2008
Dynamical systems and the transition to turbulence in linearly stable shear flows
{\it Phil. Trans. R. Soc. A} {\bf 366} 1297--1315

 Gibson J F 1999
Channelflow: a spectral NavierÐStokes simulator in C++ Technical Report Georgia Institute of Technology,  {\tt http://www.channelflow.org/}

 Hinrichsen H 2000
Non-equilibrium critical phenomena and phase transitions into absorbing states
{\it Advances in Physics\/} {\bf49} 815--958

 Lundbladh A and  Johansson A V 1991
Direct simulations of turbulent spots in plane Couette flow
{\it J.~Fluid Mech.} {\bf229} 499--516.

 Lagha M and Manneville P 2007a
Modeling transitional plane Couette flow
{\it Eur. Phys. J. B\/} {\bf 58} 433--447

 Lagha M and Manneville P 2007b
Modeling transitional plane Couette flow. Large scale flow around turbulent spots
{\it Phys. Fluids\/} {\bf 19} 094105

 Manneville P 2009
Spatiotemporal perspective on the decay of turbulence in wall-bounded flows
{\it Phys.~Rev. E\/} {\bf79} 025301 [R]; 039904 [E]

 Manneville P and Rolland J 2010
On modelling transitional turbulent flows using under-resolved direct
numerical simulations
{\it Theor. Comput. Fluid Dyn.\/} DOI 10.1007/s00162-010-0215-5

 Manneville P 2011
On the decay of turbulence in plane Couette flow
{\it Fluid Dyn. Res.} {\bf 43} 065501

 Moxey D and  Barkley D 2010
Distinct large-scale turbulent-laminar states in transitional pipe flow
{\it PNAS\/}~{\bf107} 8091--8096

  Pomeau Y 1986
Front motion, metastability and subcritical bifurcations in hydrodynamics\\
{\it Physica~D\/}~{\bf23} 3--11

 Pomeau Y 1998
Transition vers la turbulence dans les \'ecoulements parall\`eles,
Chapter~4 of Berg\'e {\it et al.} (1998)

 Prigent  A  2001
{\it La spirale turbulente: Motif de grande longueur d'onde dans les \'ecoulements cisaill\'es turbulents\/}
PhD thesis, Universit\'e Paris-Sud, Orsay,\\
{\tt http://tel.archives-ouvertes.fr/tel-00261190/en/}

\newpage

 Prigent A, Gr\'egoire G, Chat\'e H and Dauchot O 2003
Long-wavelength modulation of turbulent shear flow
{\it Physica D\/} {\bf174} 100--113

 Schumacher J and Eckhardt B 2001
Evolution of turbulent spots in a parallel shear flow
{\it Phys.~Rev.~E} {\bf 63} 046307

 Tillmark N 1995
On the streading mechanisms of a turbulent spot in plane Couette flow
{\it Europhysics Lett.} {\bf 32} 481--485

Tuckerman L S and Barkley D 2011 
Patterns and dynamics in transitional plane Couette flow
{\it Phys.~Fluids} {\bf 23} 041301

 Waleffe F 1997
On a self-sustaining process in shear flows
{\it Phys. Fluids\/} {\bf9} 883--900

}
\end{document}